# Formation and delivery of complex organic molecules

# to the Solar System and early Earth

Sun Kwok

Department of Earth, Ocean, and Atmospheric Sciences, University of British Columbia,

Vancouver, Canada

Email: skwok@eoas.ubc.ca

## 4.2.1 Introduction

Life as we know it is based on chemical building blocks of organic molecules. Although organic molecules extracted from living organisms were once believed to possess a special component called "vitality", we no longer believe "vitality" is a real physical entity – as many organic molecules extracted from living organisms can also be artificially synthesized in the laboratory. Instead, organic molecules are just molecules with the element carbon (C) and hydrogen (H) as major constituents while also containing other elements such as nitrogen (N), oxygen (O), sulfur (S), phosphorus (P). From a chemical point of view, there is no intrinsic difference between organic and other molecules.

Since the 1950s, we have learned that all chemical elements (except H and some helium, He) in our bodies originated from nucleosynthesis in interior of stars (Burbidge et al. 1957). This is the first established link between stars and life on Earth. Advances in infrared and millimetre-wave observing capabilities in the 1960s led to the discovery of molecules in space and our realization that stars can synthesize molecules and minerals during their late stages of evolution (Ziruys et al. 2016). The most surprising aspect of these discoveries was the detection of complex organics in the circumstellar envelopes of evolved stars. Infrared spectroscopic observations have shown that common, ordinary stars can synthesize complex organics during their last stage of evolution in the planetary nebulae phase, before their nuclear fuel supplies are exhausted in the following white dwarf phase. This organic synthesis takes place under an extremely low ($<10^6$ cm$^{-3}$) density environment, and occurs over very short ($10^4$ yr) time scales. These organics are ejected into the interstellar medium and quite probably have spread across the Milky Way Galaxy (Kwok 2004).

In this Chapter, we will summarize the observational evidence of stellar organic synthesis,

discuss the chemical structure of these organics, explore the possible relationship between stellar organics and those found in the Solar System, and speculate on the delivery of stellar organics to Earth and its implications on the origin of life on Earth.

## 4.2.2 Late stages of stellar evolution

Stars like our Sun maintain their luminosities by generating energy through fusing H into He in the core. After the supply of H is exhausted in the core, H is converted into He in a shell surrounding the inert He core. The outer envelope of the star expands and the luminosity increases to several hundred times the current solar luminosity ($L_\odot$). This stage is called the red giant phase. Further contraction of the core leads to the ignition of He in a shell, and He is converted into carbon (C) through the triple α reaction. The envelope of the star expands to a size more than one astronomical unit (AU, the distance between the Earth and the Sun) and the luminosity climbs to more than 3000 $L_\odot$. As the result of size expansion, the surface temperature of the star drops to 3000 K, with its emerging radiation shifting from the visible to the infrared, peaking at a wavelength of ~1 μm. This is referred to in astronomical nomenclature as the asymptotic giant branch (AGB) phase.

A strong stellar wind develops during the late AGB phase, ejecting mass in the stellar envelope into interstellar space. For stars with initial mass under 8 times the mass of the Sun, this mass loss process can remove most of the envelope mass before the ignition of C and avoids the fate of becoming a supernova. This path of evolution is followed by over 95% of all stars in our Milky Way Galaxy.

After the complete depletion of the H envelope by mass loss, a faster wind develops and compresses and accelerates the previously ejected circumstellar material into a high-density shell. As the core is gradually exposed by its diminishing envelope, the star increases its temperature from 3000 K to over 100,000 K (Figure 4.2.1). The increasing ultraviolet (UV) radiation output from the star photo-ionizes the circumstellar gas, and the resulting strong atomic emission lines create a bright optical nebula called planetary nebula (Figure 4.2.2). Planetary nebula is a short-lived phase of stellar evolution, lasting only about 20,000 years before it disperses into the interstellar medium. The hot central star (core of the progenitor AGB star) gradually burns out its H fuel and becomes a white dwarf. A detailed description of the late stages of evolution leading to the formation of planetary nebulae is given in Kwok (2000).

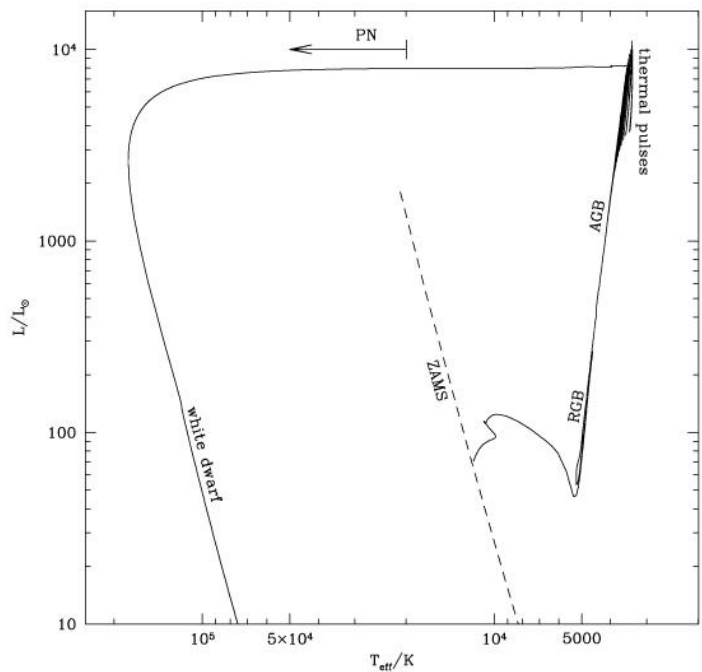

Figure 4.2.1      The evolutionary track of a 3 solar mass star on a plot of luminosity (vertical axis) vs. temperature (horizontal axis) from the zero age main sequence (ZAMS), through the red giant branch (RGB) and asymptotic giant branch (AGB) to planetary nebulae (PN) and ending as a white dwarf (Figure from T. Blöcker).    The beginning of the planetary nebulae phase is indicated by the end of an arrow, when the star reaches a temperature of ~20,000 K with sufficient output of UV photons to photoionize the surrounding nebula.    The evolutionary stage between the end of the AGB and the beginning of planetary nebulae phases is called proto-planetary nebulae phase (Kwok 1993).    During this phase, the circumstellar nebulae have no atomic line emission and are only illuminated by scattered starlight.

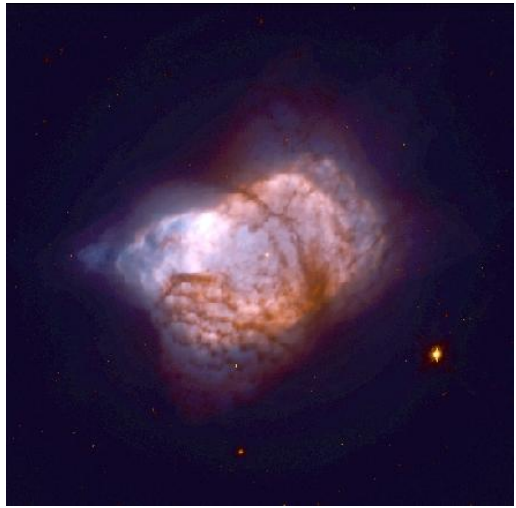

Figure 4,2,2. A composite color image of the planetary nebula NGC 7027.    The color of the nebula is due to emission lines from ions of H, O and N.    Unidentified infrared emission (UIE) bands were first discovered in the spectrum of this nebula.

### 4.2.3   Circumstellar synthesis of molecules and minerals

Carbon atoms synthesized in the core of AGB stars are dredged up to the stellar surface through the process of convection. In the low-temperature atmosphere of AGB stars, simple molecules such as $C_2$, $C_3$, and CN can form. As the envelope is ejected by the stellar wind, further chemical reactions have been observed to take place in the circumstellar envelope, leading to a rich variety of chemical species. As the gas expands and cools, atoms combine to form molecules and gas condense directly into solid phase to form micron-size solid grains. Solid particles in the stellar wind can be identified by their lattice vibrational modes in the infrared through infrared spectroscopy. For gas-phase molecular species, their rotational transitions can be detected by millimeter- and submillimeter-wave spectroscopy. As of 2018, approximately 80 molecular species have been detected in the circumstellar environment of evolved stars. The detected molecular species include inorganics (e.g., CO, SiO, SiS, $NH_3$, AlCl, etc.), organics ($C_2H_2$, $CH_4$, $H_2CO$, $CH_3CN$, etc.), radicals (CN, $C_2H$, $C_3$, $HCO^+$, etc.), chains (e.g., HCN, $HC_3N$, $HC_5N$, etc.), and rings ($C_3H_2$) (Ziurys et al. 2006, 2016).

Micron-size solid-state particles (commonly referred to as dust in the astronomical literature) can be directly detected by their thermal radiation in the infrared. The first circumstellar solid discovered was amorphous silicates, which were identified by their Si–O stretching and the Si–O–Si bending modes at 9.7 and 18 μm, respectively (Woolf and Ney 1969). These features are detected in over 4,000 O-rich AGB stars by the Low Resolution Spectrometer (LRS) on board of the *Infrared Astronomical Satellite (IRAS)* all-sky survey (Kwok et al. 1997). These features can be in emission or in self-absorption, depending on the amount of mass in the circumstellar envelope (Figure 4.2.3).

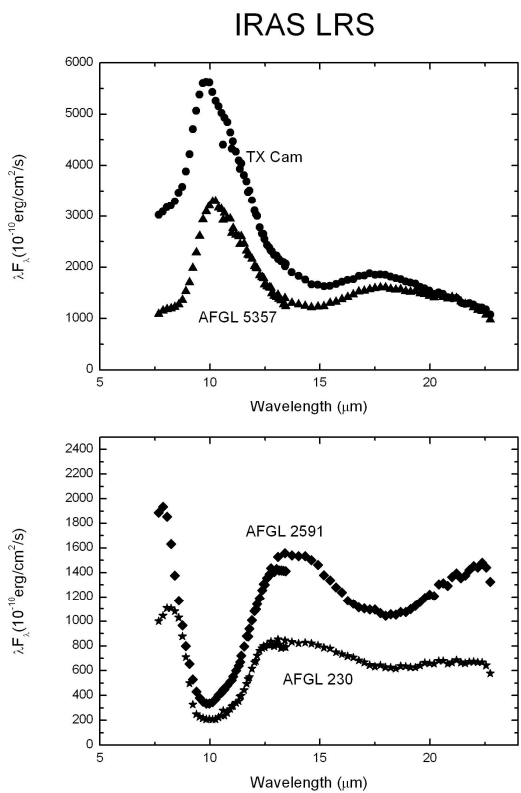

Figure 4.2.3. Infrared spectra of amorphous silicates. The 9.7 and 18 μm features of silicates can be seen both in emission (top panel) and in self absorption (bottom panel). Data obtained from the Low Resolution Spectrometer observations from the Infrared Astronomical Satellite.

A variety of refractory oxides (corundum α-$Al_2O_3$, spinel $MgAl_2O_4$, rutile $TiO_2$) are also detected in the envelopes of AGB stars (Posch et al. 1999, 2002). Crystalline silicates such as pyroxenes and olivines have sharper features are also detected in the circumstellar environment (Jäger et al. 1998). The wide detection of minerals in stars by infrared techniques led to the emergence of the field of astromineralogy (Henning 2009).

In C-rich stars where the atmospheric C abundance exceeds that of O, all O atoms are tied up in CO and the surplus C atoms form C-based molecules and solids. The most common C-based solid is silicon carbide (SiC), which 11.3 μm emission feature is observed in over 700 C-rich stars in the *IRAS LRS* survey (Kwok et al. 1997). For highly evolved carbon-rich stars, the dust component is believed to be dominated by amorphous carbon, which emits a strong featureless continuum in the infrared (Volk et al. 2000). The circumstellar dust can completely obscure the optical surface (photosphere) of the star and convert all its energy output to the infrared. These highly evolved stars have no optical counterparts and can only be detected in the infrared. They are referred to as extreme carbon stars. It is also during this very late AGB evolution stage that the molecule acetylene ($C_2H_2$) is seen (Figure 4.2.4).

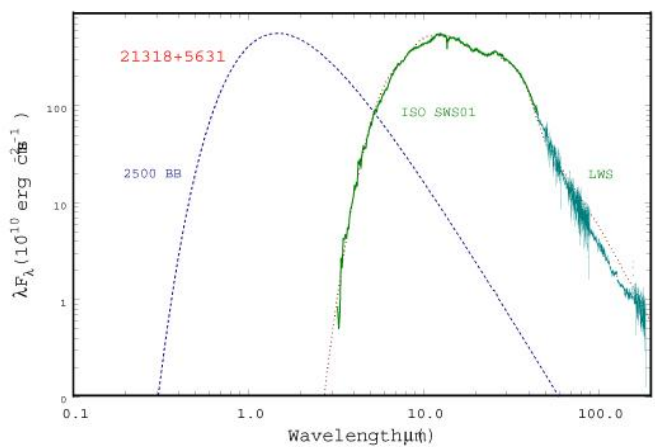

Figure 4.2.4. The extreme carbon star IRAS 21318+5631 is an example of a star so obscured by its own ejected circumstellar dust envelope that the central star becomes undetectable in the optical region. Its infrared spectrum (solid line) is completely due to dust emission and has a color temperature of 300 K. The dotted line represents the theoretical fit to the spectrum based on a 1-D radiation transfer model with a hidden 2500 K central star (dashed line) as the energy source. The absorption feature near the peak of the spectrum is the 13.7 μm band of acetylene.

In the terrestrial environment, formation of molecules and solids are the results of 3- or multi-body collisions. The densities of the stellar winds, however, are much too low ($<10^6$ H atoms per c.c.) for such processes. From observations of novae, we learn that silicate dust condenses as soon as the radiation temperature drops below the condensation temperature, regardless of the density conditions. The emergence of a dust spectrum from a pure gas spectrum in novae can take place over a time span as short as a few days (Ney and Hatfield 1978). From mapping of molecular distribution in the circumstellar envelopes of AGB stars by millimeter-wave interferometric techniques, we can set limits to the formation time of circumstellar molecules as hundreds of years based on the sizes of the molecular emitting regions and expansion velocities of the envelopes. These observations suggest that molecule and solid synthesis in the circumstellar environment can be extremely efficient, in spite of the low-density environment.

Since the initial discovery of fullerenes in the laboratory (Kroto et al. 1985), there have been strong interests in searching for this molecule in space. $C_{60}$ is now unambiguously detected in planetary nebulae (Cami et al. 2010, García-Hernández et al. 2010), reflection nebulae (Sellgren et al. 2010) and in proto-planetary nebulae (Zhang and Kwok 2011, 2013). These cage-like molecules are detected either through the C–C or C–H stretching vibrational modes in the infrared. With 60 or more C atoms, fullerenes are the heaviest molecular species detected in the envelopes of stars. The paths of synthesis of $C_{60}$ have been suggested to be either top down (as breakdown products of complex organics) or bottom-up (built up from small C-based molecules) (García-Hernández et al. 2012, Bernard-Salas et al. 2012). As broad plateau emission features always accompany $C_{60}$ features (Figure 4.2.5), it is possible that $C_{60}$ synthesis is related to the unidentified infrared emission (UIE) phenomenon (see next section).

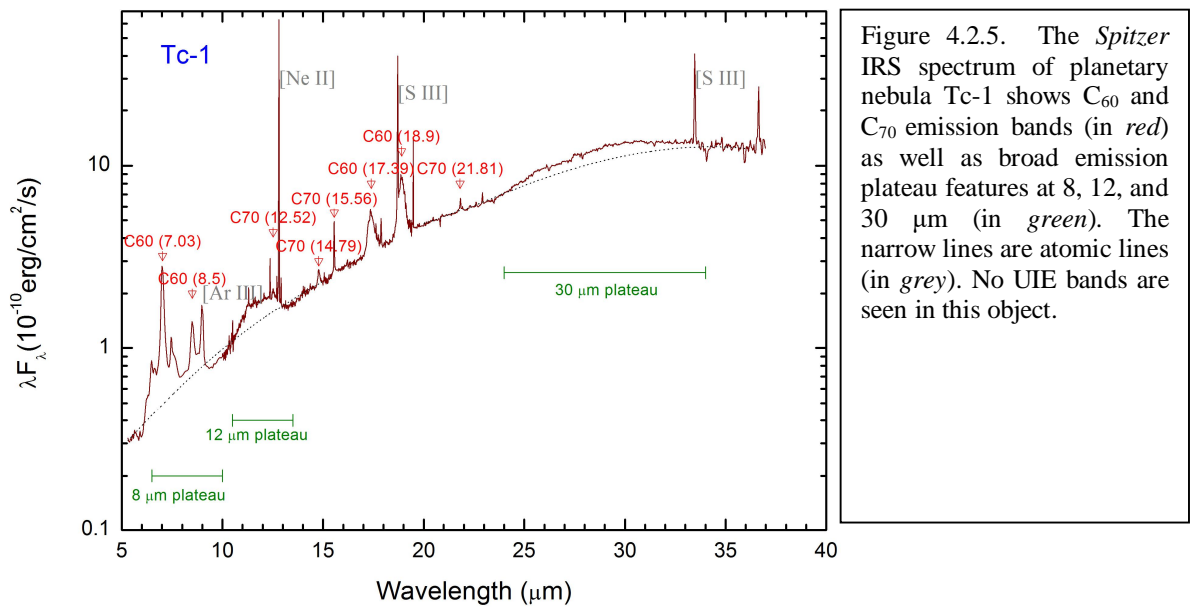

Figure 4.2.5. The *Spitzer* IRS spectrum of planetary nebula Tc-1 shows $C_{60}$ and $C_{70}$ emission bands (in *red*) as well as broad emission plateau features at 8, 12, and 30 μm (in *green*). The narrow lines are atomic lines (in *grey*). No UIE bands are seen in this object.

In a circumstellar environment, H can easily attach to the fullerene molecule and become hydrogenated fullerenes (or fulleranes, $C_{60}H_m$, $m=1–60$). The theoretical vibrational spectra of fulleraness can be calculated and compared to observed astronomical spectra, resulting in the suggestion that fulleranes may be common in the circumstellar environment (Zhang and Kwok 2013, Zhang et al. 2017).

### 4.2.4   Unidentified infrared emission bands

Due to the low-density and high UV radiation background of the interstellar medium, complex organic molecules and solids were not believed to be able to exist in space. The detection of the unidentified infrared emission (UIE) features in the spectra of planetary nebulae, now attributed to organics, came as a complete surprise. A family of broad infrared bands at 3.3, 6.2, 7.7, 8.6 and 11.3 μm was first found in the spectrum of the planetary nebula NGC 7027 (Russell et al. 1977) (Figure 4.2.2 and 4.2.6). Shortly after discovery, it was recognized that the UIE bands probably arise from the vibrational modes of organic compounds (Knacke 1977, Duley and Williams 1979). More specifically, the UIE features are suggested to be originating from stretching and bending vibrational modes of aromatic compounds (Duley and Williams 1981). Also present in spectra are emission features around 3.4 μm, which arise from symmetric and anti-symmetric C–H stretching modes of methyl and methylene groups (Puetter et al. 1979, Geballe et al. 1992). The bending modes of these groups also manifest themselves at 6.9 and 7.3 μm (de Muizon et al. 1990, Chiar et al. 2000). In addition, there are weaker unidentified emission features at 15.8, 16.4, 17.4, 17.8, and 18.9 μm which probably arise from C skeleton vibrational modes.

The emission bands themselves are often accompanied by strong, broad emission plateaus features at 6–9, 10–15, and 15–20 μm. The first two plateau features have been identified as superpositions of in-plane and out-of-plane bending modes emitted by a mixture of aliphatic side groups attached to aromatic rings (Kwok et al. 2001).

The UIE bands are extremely prominent in the spectra of C-rich planetary nebulae (Figure 6). They are not seen in AGB stars but emerge during the proto-planetary nebula phase. Their existence suggests that organic compounds can be efficiently synthesized in the circumstellar environment over very short ($10^3$ yr) time scales.

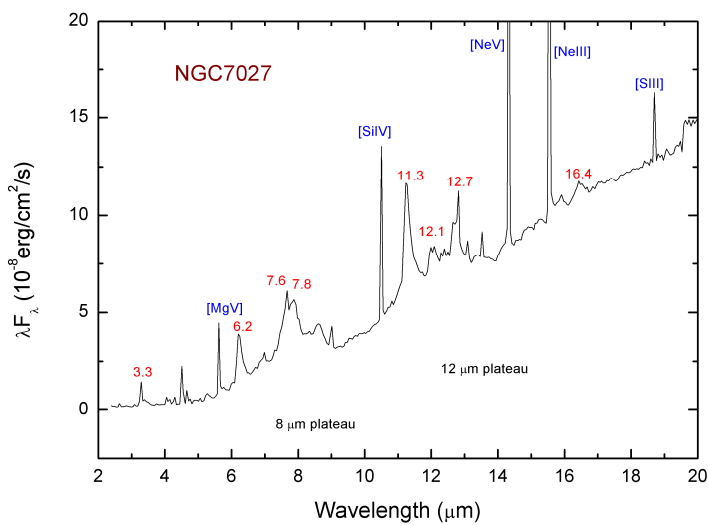

Figure 4.2.6. *Infrared Space Observatory (ISO) spectrum of the planetary nebula NGC 7027 showing the UIE bands (labeled in red with the wavelength of peak emission in units of microns). Broad emission plateaus around 8 and 12 μm as well as a strong underlying continuum can be seen. The narrow lines (in blue) are atomic lines.*

The UIE and the plateau emission features lie on top of a strong continuum that extends from the near infrared to millimeter wavelengths. This continuum emission must be due to thermal emission from micron-size solid grains, similar to those observed in AGB stars (Figure 4.2.4). In planetary nebulae or active galaxies, most of the output energy of the object is emitted through this continuum component (Figure 4.2.7). Even in the diffuse interstellar medium, the strength of the UIE features are strongly correlated with the dust continuum, suggesting that the heating source for the UIE carriers and the dust continuum must be the same (Kahanpää et al. 2003).

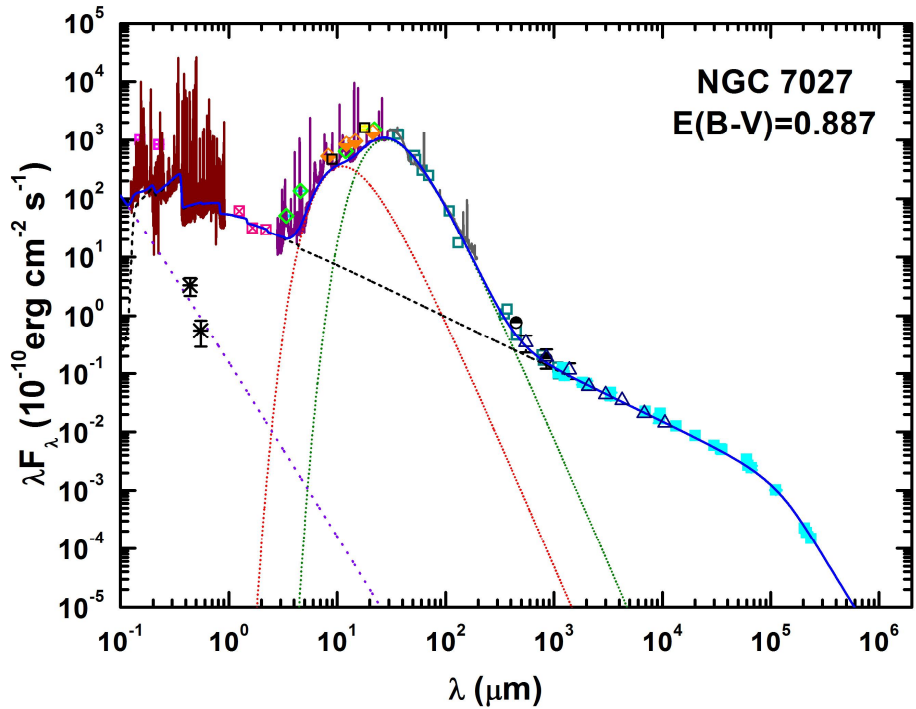

Figure 7. The spectral energy distribution of the planetary nebula NGC 7027 from wavelength 0.1 μm (ultraviolet) to 10 cm (radio). The continuum emission of the object is shown in a red line. The continuum shortward of wavelengths of 3 μm is due to bound-free gas emission. Between wavelengths 1 μm and 1 mm, the continuum is due to solid-state particle (dust) emission. At wavelengths long ward of 1 mm, free-free gas emission dominates. We can see that most of the energy output is emitted by a dust component which is approximated by a sum of two blackbodies (dotted lines) of 246 and 97 K. The narrow lines above the continuum are atomic lines and UIE bands.

### 4.2.5 Carrier of the UIE bands

Noting that the astronomical UIE spectra resemble the spectra of automobile exhaust soot particles, Allamandola et al. (1985) suggested that the UIE carriers are simple polycyclic aromatic hydrocarbon (PAH) molecules. It is now widely believed in the astronomical community that the UIE features are due to infrared fluorescence of far-UV-pumped PAH molecules each containing ~50 C atoms (Tielens 2008). These small, gas-phase PAH molecules can be excited by a single incoming UV photon stochastically to a high temperature over a short time interval, therefore accounting for the strengths of the UIE features under low-temperature interstellar conditions (Sellgren 1984). By engaging a mixture of hundreds of PAH molecules of different sizes, structures (compact, linear, branched) and ionization states, and broadened by artificial broad intrinsic line profiles, the PAH model can fit the astronomical UIE spectra.

The fitting of the astronomical UIE features by PAH molecules is not straight forward. Although PAH molecules were known to have C–H stretching modes around 3.3 μm, the actual wavelengths of the C–H stretching is short ward of the 3.3 μm UIE band (Sakata et al. 1990, Kwok and Zhang 2013). PAH molecules generally have out-of-plane bending modes around 10-14 μm, but to fit the observed UIE 11.3 μm feature requires exotic mix of PAH molecules (Sadjadi et al. 2015b). The assignments of the 6.2, 7.7, and 8.6 μm UIE features to PAH molecules are even more difficult. It was suggested that the 6.2 μm feature is due to C–C stretching and the 8.6 μm feature is due to C–H in-plane bending modes of PAH molecules (Léger and Puget 1984). The 7.7 μm UIE feature has no obvious counterpart in PAH spectra and has been suggested to be due to blending of several C–C stretching modes (Allamandola et al. 1989). Since C–C stretching modes in PAH molecules are generally very weak, it was later proposed that the 6.2 and 7.7 μm bands are due to ionized PAH molecules, which tend to show stronger C–C stretching modes (Hudgins and Allamandola 1999, van Diedenhoven et al. 2004). However, experimental results show that the C–C vibrational modes of PAH ions generally occur at wavelengths longer than 6.2 μm (Bauschlicher 2002, Hudgins et al. 2005). By increasing the size of the molecules, the peak of the band can shift to shorter wavelengths, but never as short as 6.2 μm (Hudgins et al. 2005). In order to account for this discrepancy, it was suggested that some of the C atoms in specific positions in the ring be replaced by nitrogen (N) and this modified PAH model is referred to as the PANH model.

In addition, the PAH hypothesis suffers from the following problems: (i) PAH molecules have well-defined sharp features but the UIE features are broad; (ii) PAHs are primarily excited by UV, with little absorption in the visible, but UIE features are seen in proto-planetary nebulae and reflection nebulae, objects with very little UV background radiation and the shapes and peak wavelengths of UIE features are independent of temperature of the exciting stars (Uchida et al. 2000); (iii) the strong and narrow PAH gas phase features in the UV are not seen in interstellar extinction curves to very low upper limits (Clayton et al. 2003, Salama et al. 2011, Gredel et al. 2011); (iv) no specific PAH molecule has been detected in spite of the fact that the vibrational and rotational frequencies of PAH molecules are well known; (v) there are great difficulties in reconciling the with band positions or relative intensities of laboratory PAH spectra with astronomical UIE spectra (Wagner et al. 2000); (vi) the large number of free parameters in the PAH model fitting suggests that such fittings are not very meaningful (Zhang and Kwok 2015).

In response to these criticisms, the PAH model has been revised to incorporate ionization states and large sizes to increase the absorption cross sections in the visible, introduce dehydrogenation, superhydrogenation and minor aliphatic side groups to explain the aliphatic features, appeal to a large mixture of different PAH molecules to explain the lack of detection of individual PAH molecules. Since known PAH molecules have problems reproducing the wavelengths of the UIE bands, a large mixture of diverse PAH molecules is needed to fit the observed astronomical spectra. Hetro atoms such as N and O are also introduced to explain the 6.2 and 11.3 μm features. The PAH hypothesis therefore has moved away from the chemical definition of PAH molecules to a hybrid to save the hypothesis.

In order to identify possible carriers of the UIE bands, it would be useful to see what carbonaceous products can naturally exist in the interstellar medium. By introducing H into graphite and diamond (both crystalline forms of pure carbon), a variety of amorphous C–H alloys can be created (Robertson and O'Reilly 1987, Jones et al. 1990, Jones 2012a, b, c, Jones et al. 2013). Different geometric structures with long- and short-range can be created by varying the aromatic to aliphatic and C to H ratios. The infrared spectra of these amorphous carbonaceous materials (Dischler et al. 1983) show resemblance to the astronomical UIE bands seen in planetary nebulae and proto-planetary nebulae (Figure 8). Since these amorphous carbonaceous solids have absorption bands in the visible, they can be easily excited by visible light from stars.

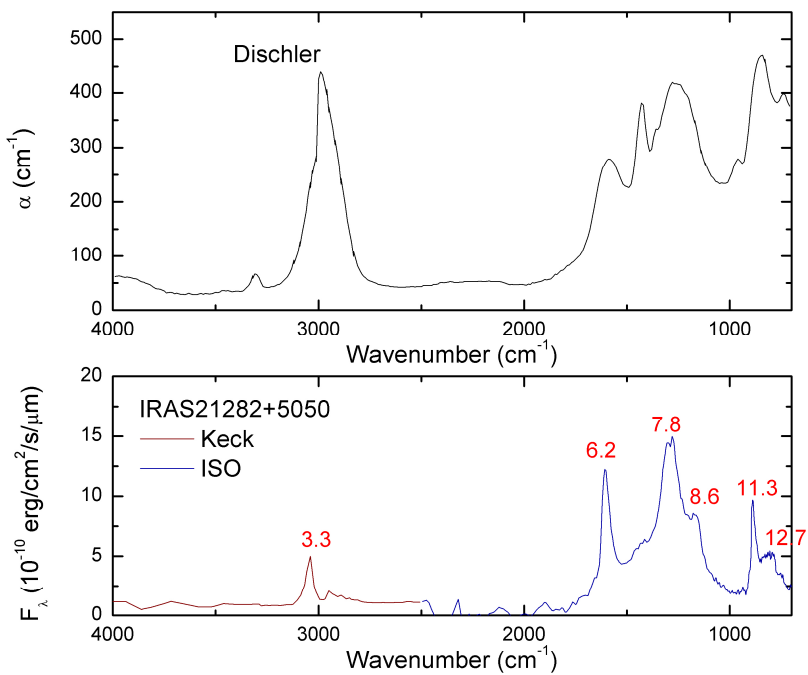

Figure 4.2.8. Laboratory infrared spectra (inverted from absorption) of hydrogenated amorphous carbon (top, from Dischler et al. 1983) compared to the astronomical spectrum of the planetary nebula IRAS 21282+5050 (bottom panel). The UIE bands are labeled by their wavelengths in μm.

By the early 1980s, it was known that carbon clusters can be produced by laser vaporization of graphite followed by supersonic expansion into an inert gas. The employment of this technique has led to the discovery of fullerene, a new form of carbon (Kroto et al.1985). Since then, various techniques based on laser pyrolysis of gas-phase hydrocarbons followed by condensation have been used to create laboratory counterparts of cosmic organic dust (Jäger et al. 2009). These include the quenching of plasma of 4-torr methane (Sakata et al. 1987), hydrocarbon flame or arc-discharge in a neutral of hydrogenated atmosphere (Colangeli et al. 2003, Mennella et al. 2003), laser ablation of graphite in a hydrogen atmosphere (Scott and Duley 1996, Mennella et al. 1999), infrared laser pyrolysis of gas phase hydrocarbon molecules (Herlin et al. 1998), and photolysis of methane at low temperatures (Dartois et al. 2004).

Amorphous carbonaceous solids are known to be naturally produced through the process of combustion. Soot is formed by igniting a mixture of gas-phase hydrocarbons with oxygen, resulting in amorphous structures consisting of islands of aromatic rings linked by aliphatic chains (Pino et al. 2008).

In addition to soot, there are also other natural decayed products of living organisms such as coal and kerogen that have similar amorphous mixed aromatic/aliphatic properties (Painter et al. 1981). The infrared spectra of soot (Keifer et al. 1981), coal (Ibarra et al. 1996, Guillois et al. 1996, Papoular 2001) and petroleum and asphaltenes (Cataldo et al. 2002, Cataldo et al. 2013) all show spectral features similar to the astronomical UIE bands.

As the results of these laboratory developments, alternate models for the UIE phenomenon have been proposed. These include hydrogenated amorphous carbon (HAC, Duley 1993), soot and carbon nanoparticles (Hu and Duley 2008), quenched carbonaceous composite particles (QCC, Sakata et al. 1987), kerogen and coal (Papoular et al. 1989), petroleum fractions (Cataldo et al. 2002), and mixed aromatic/aliphatic organic nanoparticles (MAON, Kwok and Zhang 2011, 2013). In the coal, petroleum, and MAON models, other elements such as O, S, N are also incorporated into the hydrocarbon compounds. In a natural environment such as circumstellar envelopes of evolved stars where a mix cosmic elements are present, it is expected any organic product of synthesis will contain other elements beyond C and H. A schematic illustration of part of a MAON structure is shown in Figure 4.2.9.

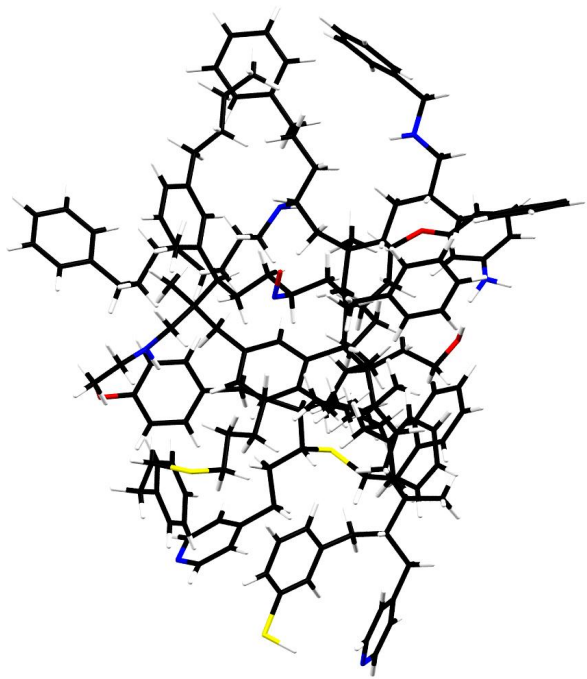

Figure 9. A 3-D illustration of a possible partial structure of a MAON particle. Carbon atoms are represented in black, hydrogen in light grey, sulphur in yellow, oxygen in red, and nitrogen in blue. There are 169 C, 225 H, 7 N, 4 O, and 3 S atoms in this example.

The advantages of such amorphous models are that the vibrational bands are naturally broad and do not needed to be artificially broadened as in the case of PAH molecules. The disadvantages are that although some experimental spectra of these materials have been obtained in the laboratory, the exact vibrational modes of the features are not known. Theoretical calculations are needed to analyze the vibrational modes of HAC, QCC, and MAONs (Sadjadi et al. 2015a). If these molecules are too large to be transiently heated by single photon excitation, alternative excitation mechanism such as chemical excitation needs to be considered (Duley and Williams 1988, 2011).

### 4.2.6 Other unidentified circumstellar spectral phenomena

There are a number of unidentified circumstellar spectral phenomena which carriers are likely to be organic compounds. The 30 μm feature was discovered in C-rich AGB stars and planetary nebulae (Forrest et al. 1981) and the 21μm feature was discovered in proto-planetary nebulae (Kwok, Volk and Hrivnak 1989). High resolution *ISO* and *Spitzer* observations have found that all observed 21μm features have the same intrinsic profile and peak wavelength (20.1 μm) (Volk, Kwok and Hrivnak 1999, Hrivnak, Volk and Kwok 2009). An example of the 21 and 30 µm features is shown in Figure 4.2.10. Strong 8 and 12 µm plateau features are seen, suggesting that there is a link between the 21 and 30 µm features and the UIE phenomenon.

These two unidentified emission features can carry a large fraction of the total energy output of the central stars – up to 8% for the 21 µm and 20% for the 30 µm features (Hrivnak et al. 2000). This implies that carrier must be made of common elements. The fact that the features are seen only in C-rich (based on photospheric absorption spectrum) objects suggests that the carrier is carbonaceous.

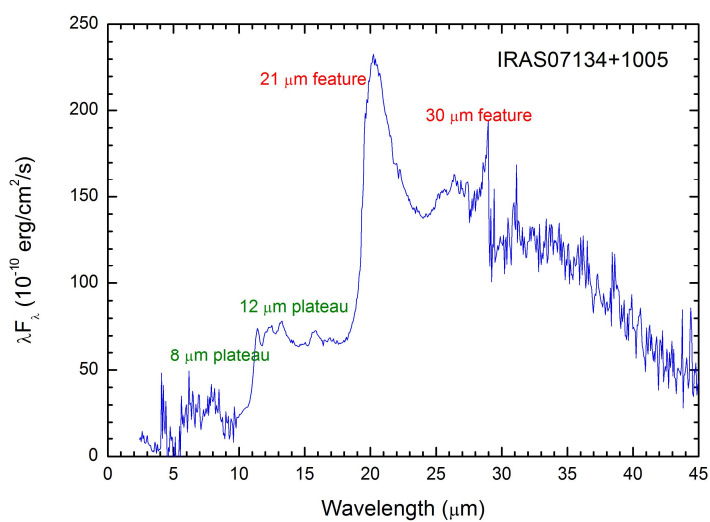

Figure 10        Combined *ISO* SWS and LWS and *Spitzer* IRS spectra of the proto-planetary nebula IRAS07134+1005 showing the 21 and 30 unidentified emission features. No atomic line is seen in the infrared spectrum of this proto-planetary nebula. Also seen in the spectra are broad plateau emission features around 8 and 12 µm. All these features sit on top of a strong continuum.

The extended red emission (ERE) is a broad band optical emission seen in reflection nebulae and planetary nebulae. It is likely to be the result of photoluminescence from semi-conductor solid particles (Witt et al. 1998). Some of the over 500 unidentified diffuse interstellar bands (DIB, Sarre 2006) observed in absorption in the diffuse interstellar medium are also seen in circumstellar envelopes (Van Winckel et al. 2002), suggesting that some of the carriers of DIBs are also synthesized in circumstellar envelopes.

### 4.2.7    Time sequence of circumstellar organic synthesis

The detection of simple molecules in early AGB stars to complex organics in planetary nebulae suggests that the synthesis of organic dust is not a breakdown of biological compounds but a bottom-up synthesis from simple molecules. The formation of the linear molecule acetylene in highly evolved AGB stars (Volk et al. 2000) is the first step for the formation of benzene (Cernicharo et al. 2001), the simplest aromatic molecule, in proto-planetary nebulae. The emergence of aromatic and aliphatic bands in the transition from AGB to planetary nebulae suggests that small rings group into larger aromatic islands, attached with aliphatic side groups. Since circumstellar chemical synthesis must occur over times scales much shorter than the dynamical time scale of expansion of the nebula and evolution time scales of the central star, we know that circumstellar organic synthesis occurs at $10^3$-$10^4$ yr time scales.

### 4.2.8 Relationship between stellar, interstellar and Solar System organics

Although it was commonly believed that Solar System objects were made of minerals, metals, and ices, we now know that meteorites, asteroids, comets, planetary satellites, and interplanetary dust particles all contain organic materials. Since the early discoveries of hydrocarbons and amino acids in meteorites (Nagy et al. 1961, Kvenvolden et al. 1970), we now know that meteorites contain rich assortment of prebiotic compounds which are entirely of abiotic origin. In the soluble component of carbonaceous chondrites, almost all biologically relevant organic compounds can be found (Schmitt-Kopplin et al. 2010).

The insoluble organic matter (IOM) of carbonaceous chondrites is composed of highly substituted single ring aromatics, substituted furan/pyran functional groups, highly branched oxygenated aliphatics and carbonyl groups (Cody et al. 2011). The 3.4 μm feature found in proto-planetary nebulae, planetary nebulae, and in the diffuse interstellar medium, is also detected in meteorites (Cronin and Pizzarello1990, Ehrenfreund et al. 1991), interplanetary dust particles (Flynn et al. 2003), comets (Keller et al. 2006), and in the haze of Titan (Kim et al. 2011) and Saturn (Kim et al. 2012). The relative elemental abundance in the IOM ranges from $C_{100}H_{70}N_3O_{12}S_2$ for Murchison meteorite to $C_{100}H_{46}N_{10}O_{15}S_{4.5}$ for the Tagish Lake meteorite (Pizzarello and Shock 2017). The detection of anomalous isotopic ratios in elements in IOM suggests that at least part of the IOM is of presolar origin.

### 4.2.9 Delivery of stellar organics to the Solar System

Since most (~95%) stars in our Galaxy go through the AGB and planetary nebulae phases of evolution, large amounts of mineral and organic solids are produced by ordinary stars over the last ten billion years in our Milky Way Galaxy. These products are ejected into the interstellar medium and spread throughout the Galaxy and evolved stars therefore represent a major source of molecules and solids in the Galaxy. Direct evidence for stellar grains to have traveled across the Galaxy to arrive at the Solar System and to Earth can be found in the form of pre-solar grains in meteorites. Pre-solar grains are grains of diamonds, silicon carbide, corundum, and spinel that have isotopic abundances not typical of the Solar System but consistent with origin in AGB stars (Zinner, 1998, Davis 2011). Since macroscopic organics such as MAON, kerogen and IOM are extremely sturdy, they are therefore likely to be able to transverse across interstellar space even under high UV radiation background and dynamical shock conditions. Stardust therefore plays an important role in the chemical enrichment of the Galaxy, and possibly also the primordial solar system (Kwok, 2004; Ziurys et al., 2016).

Possible links between Solar System organics and stellar organics can be tested by isotopic ratios. The isotopic signatures of IOM in the Murchison meteorite are consistent with at least in part being presolar in origin (Pizzarello and Shock 2017). Further work is needed to determine the origin of Solar System organics, whether these organics were made in the early Solar System, or were delivered from stellar sources

## 4.2.10 Organic reservoir on Earth

The Earth has a large reservoir of organics, almost all of which are biological in origin. The total amount of C in the living biosphere (Earth's crust, ocean and atmosphere) is about 1000 Gigatons (GT). In addition, there are about 4000 GT of fossil fuels in the form of petroleum, coal, and natural gases. The majority of which –15,000,000 GT– are in the form of kerogen, a macromolecular compound found in sedimentary rocks (Falkowski et al. 2000). Although kerogen and fossil fuels are all remnants of past life, there is a small amount of hydrocarbons found in hydrothermal vents which are of abiological origin (Sherwood Lollar et al. 2002).

The possibility that there may be primordial organics in Earth was suggested by Gold (1999). At that time of Gold's writing, gas-phase organic molecules have already been known to be widely present in interstellar space. Gold suggested that the early Earth may have trapped inside a large reservoir of methane, which could serve as precursors of fossil fuels which later flows from the interior to the surface. This theory was never taken seriously by geologists because interstellar methane gas would have great difficulty surviving the high temperature and shock during the coalescence process of Earth formation.

At a more general level, although abiotic theories of the origin of oil were seriously discussed in the Soviet Union, they were never popular in the west, in part because of the well-established links between petroleum and life in the form of nickel- and vanadium-porphyrin complexes in petroleum and iron-porphyrin and hemoglobin in animals and magnesium chlorophyll in plants. However, in addition to biological formation of fossil fuels, there exists the additional possibility of primordial organics deep inside the Earth. If the primordial hydrocarbon was in the form of a macromolecular compound, its chances of survival during Earth's formation would have been much higher that the case for methane. If such primordial macro-organics are embedded deep inside the Earth, they would be difficult for us to discover with present technology. If such primordial organics does exist and can be retrieved, the economic impact could be immense.

## 4.2.11 Effects of star dust on the origin of life on Earth

Since the Oparin-Haldane hypothesis (Oparin 1938, Haldane 1929) and the Miller-Urey experiment (Miller 1953), it has been commonly assumed that life on Earth originated from chemical reactions of simple ingredients in a suitable environment (primordial soup). The theory endogenous synthesis as the basis of origin of life has been the dominant theory in the past 50 years. However, with the observation of stellar synthesis of organics and the wide presence of complex organics among Solar System objects, the exogenous delivery hypothesis has been gaining acceptance. The most widely discussed agents for exogenous delivery are comets, asteroids, interplanetary dust particles and micrometeorites. The influx rate at present for micrometeorites is estimated to be ~30,000 tons $yr^{-1}$, but the rates were much higher during the early 500 million years of the Earth's history during the period of heavy early bombardment. Stellar MAON-like materials embedded in comets and asteroids have a good chance of survival upon impact.

We cannot rule out the possibility that the planetismals that aggregated to form the primordial

Earth may have contained remnants of macromolecular star dust. These organics could well survive the heat and shock conditions during the Earth-formation process. If this is the case, then there could exist primordial organics deep inside the interior of the Earth (Kwok 2017).

Under suitable temperature and pressure conditions, e.g. those in hydrothermal vents, organic components, such as hydrocarbons, dicarboxylic acids, N-, O-, and S-containing aromatic compounds, as well as ammonia can be released from macromolecular compounds (Yabuta et al. 2007, Pizzarello et al. 2011). These prebiotic materials could form the ingredients of the first steps to life in the early Earth. Whether life on Earth originated on the surface in water, or deep inside the Earth, is an interesting question for further exploration.

### 4.2.12 Conclusions

In the early 20[th] century, the discipline of astrophysics was born out of development of atomic and nuclear physics. The discipline of astrochemistry, or the study of molecules and minerals in space, began only 50 years ago as the result of advances in observing capabilities in the infrared and millimeter-wave regions of the electromagnetic spectrum. The discipline of astrobiology, trying to bridge the gap between molecules and life, is still in its infancy. The unexpected discovery of common presence of organics in the Solar System and in the interstellar medium gives us the optimism and hope for the future of this new discipline.

From the analysis of organic matter in meteorites and other Solar System objects, we now know that abiotic synthesis can create a wide range of organic compounds far beyond those found on Earth. This hints that the degree of complexity and diversity of prebiotic organics is much larger than previously thought (Meringer and Cleaves 2017). It is therefore not surprising that there still remains a number of unexplained astronomical spectral phenomena, as the carrier of these phenomena may be organic compounds that we are unfamiliar with in the terrestrial environment.

The circumstellar envelopes of evolved stars provide the only laboratory where such abiotic organic synthesis is directly observed to take place. Based on spectroscopic observations of evolved stars in different stages of evolution, we learn that organic molecules and solids can be synthesized in near-vacuum conditions over very short time scales. This suggests that our understanding of chemical reactions based on terrestrial conditions is inadequate to explain chemistry at work in the Universe.

From the fact that UIE features are seen in galaxies with redshifts as high as 2, we know that organic synthesis took place soon after the nucleosynthesis of the element carbon, as early as 10 billion years ago (Kwok 2011). Since our Sun and our Solar System only came into existence 4.6 billion years ago, life could have emerged a very long time ago elsewhere in our Galaxy and beyond. Since terrestrial biochemistry only represents a very small branch of the possible biochemical pathways in the Universe, life as we know it can be totally different from life evolved from similar prebiotic ingredients but under different conditions and pathways. If some of these extraterrestrial life had developed into intelligent life, the level of extraterrestrial intelligence is far beyond our imagination. Our present study of the abiotic synthesis of complex organics in the circumstellar environment therefore represents a small, but important step in our understanding of intelligent life in the Universe.

Acknowledgements
SK thanks Yong Zhang, Chih-Hao Hsia, and SeyedAbdolreza Sadjadi for helpful discussions. This work is supported by a grant from the Natural Sciences and Engineering Research Council of Canada.

**References**

Allamandola, L.J., A. G. G. M. Tielens, and J.R. Barker. 1985. Polycyclic aromatic hydrocarbons and the unidentified infrared emission bands - auto exhaust along the Milky Way. Astrophys. J. 290: L25-L28.

Allamandola, L.J., A.G.G.M. Tielens, and J.R. Barker. 1989. Interstellar polycyclic aromatic hydrocarbons: the infrared emission bands, the excitation / emission mechanism and the astrophysical implications. Astrophys. J. Suppl. Ser. 71:733-775.

Bauschlicher, C.W., Jr. 2002. The infrared spectra of $C_{96}H_{24}$, $C_{96}H^+_{24}$, and $C_{96}H^+_{25}$. Astrophys. J. 564:782-786.

Bernard-Salas J, J. Cami, E. Peeters, et al. 2012. On the excitation and formation of circumstellar fullerenes. Astrophys. J. 757:41.

Burbidge, E.M., G.R., Burbidge, W.A. Fowler, and F. Hoyle. 1957. Synthesis of the Elements in Stars. Rev. Mod. Phys. 29, 547 – 650.

Cami, J., J. Bernard-Salas, E. Peeters, and S.E. Malek. 2010. Detection of $C_{60}$ and $C_{70}$ in a young planetary nebula. Science 329:1180-1182.

Cataldo, F., Y. Keheyan, and D. Heymann. 2002. A new model for the interpretation of the unidentified infrared bands (UIBS) of the diffuse interstellar medium and of the protoplanetary nebulae. International J. Astrobiology 1:79-86.

Cataldo, F, D.A. García-Hernández, and A. Manchado. 2013. Far- and mid-infrared spectroscopy of complex organic matter of astrochemical interest: coal, heavy petroleum fractions and asphaltenes. Mon. Not. R. Astron. Soc. 429:3025-3039.

Cernicharo J, A.M. Heras, A.G.G.M. Tielens et al. 2001. Infrared Space Observatory's discovery of $C_4H_2$, $C_6H_2$, and benzene in CRL 618. Astrophys. J. 546(2): L123-L126.

Chiar, J.E., A.G.G.M. Tielens, D.C.B. Whittet et al. 2000. The composition and distribution of dust along the line of sight toward the Galactic Center. Astrophys. J. 537:749-762.

Clayton, G.C., K.D. Gordon, F. Salama et al. 2003. The role of polycyclic aromatic hydrocarbons in ultraviolet extinction. I. probing small molecular polycyclic aromatic hydrocarbons. Astrophys. J. 592:947-952.

Cody, G.D., E. Heying, C.M.O. Alexander, et al. 2011. Establishing a molecular relationship between chrondritic and cometary organic solids. Proceedings of the National Academy of Science 108 (48):19171-19176.

Colangeli, L.,Th. Henning, J.R. Brucato et al. 2003. The role of laboratory experiments in the characterisation of silicon-based cosmic material. Astron. & Astrophys. Rev. 11(2-3):97-152

Cronin, J.R., and S. Pizzarello. 1990. Aliphatic hydrocarbons of the Murchison meteorite. Geochimica et Cosmochimica Acta, 54, 2859-2868.

Dartois, E., G.M. Muñoz Caro, D. Deboffle, and L. d'Hendecourt. 2004. Diffuse interstellar medium organic polymers: photo-production of the 3.4, 6.85 and 7.25 μm features. Astron. Astrophys. 423:L33-L36.

Davis, A.M. 2011. Cosmochemistry special feature: stardust in meteorites. Proceedings of the National Academy of Science 108:19142-19146.

De Muizon, M.J., L.B. d'Hendecourt, and T.R. Geballe. 1990. Polycyclic aromatic hydrocarbons in the near-infrared spectra of 24 IRAS sources. Astron. Astrophys. 227(2):526-541.

Dischler, B., A. Bubenzer, and P. Koidl. 1983. Bonding in hydrogenated hard carbon studied by optical spectroscopy. Solid State Commun. 48(2):105-108.

Duley, W.W. 1993. Infrared spectra of interstellar carbon solids. In Astronomical Infrared Spectroscopy: Future Observational Directions, ed. S. Kwok, ASP Conference Series, p. 241.

Duley, W.W. and D.A. Williams. 1979. Are there organic grains in the interstellar medium? Nature 277:40-41.

Duley, W.W. and D.A. Williams. 1981. The infrared spectrum of interstellar dust - Surface functional groups on carbon. Mon. Not. R. Astron. Soc. 196:269-274.

Duley, W.W. and D.A. Williams. 1988. Excess infrared emission from large interstellar carbon grains. Mon. Not. R. Astron. Soc. 231:969-975.

Duley, W.W. and D.A. Williams. 2011. Excitation of the aromatic infrared emission bands: chemical energy in hydrogenated Amorphous Carbon Particles? Astrophys. J. Let. 737: L44.

Ehrenfreund, P., F. Robert, L. d'Hendecourt, and F. Behar. 1991. Comparison of interstellar and meteoritic organic matter at 3.4 microns. Astron. Astrophys., 252, 712-717.

Falkowski, P., R.J. Scholes, E. Boyle et. al. 2000. The global carbon cycle: a test of our knowledge of Earth as a system, Science 290, 291-296.

Flynn, G. J., L.P. Keller, M. Feser, S. Wirick, and C. Jacobsen. 2003. The origin of organic matter in the solar system: evidence from the interplanetary dust particles. Geochimica et Cosmochimica Acta, 67, 4791-4806.

Forrest, W.J., J.R. Houck, and J.F. McCarthy. 1981. A far-infrared emission feature in carbon-rich stars and planetary nebulae. Astrophys. J. 248:195-200.

García-Hernández, D.A., A. Manchado, P. Garcia-Lario et al. 2010. Formation of fullerenes in H-containing planetary nebulae. Astrophys. J. 724: L39-L43.

García-Hernández D.A., E. Villaver, P. Garcia-Lario et al. 2012. Infrared study of fullerene planetary nebulae. Astrophys. J. 760:107.

Geballe, T.R., A.G.G.M. Tielens, S. Kwok, and B.J. Hrivnak. 1992. Unusual 3 micron emission features in three proto-planetary nebulae. Astrophys. J. 387: L89-L91.

Gredel R, Y. Carpentier, G. Rouillé, M. Steglich, F. Huisken, and Th. Henning. 2011. Abundances of PAHs in the ISM: confronting observations with experimental results. Astron. Astrophys. 530: 26.

Guillois, O., I. Nenner, R. Papoular, and C. Reynaud. 1996. Coal models for the infrared emission spectra of proto--planetary nebulae. Astrophys. J. 464:810-817.

Gold, T. 1999. The Deep Hot Biosphere. Copernicus Books, New York.

Haldane, J.B.S. 1929. The Origin of Life, Rationalist Annu. 148, 3–10.

Henning, T. 2009. Astromineralogy. Springer.

Herlin N, I. Bohn, C. Reynaud, M. Cauchetier, A. Galvez, and J.-N. Rouzaud. 1998. Nanoparticles produced by laser pyrolysis of hydrocarbons: analogy with carbon cosmic dust. Astron. Astrophys. 330:1127-1135.

Hrivnak, B.J., K. Volk, and S. Kwok. 2009. A Spitzer study of 21 and 30 μm emission in several galactic carbon-rich protoplanetary nebulae. Astrophys. J. 694:1147-1160.

Hrivnak, B.J., K. Volk, and S. Kwok. 2000. 2-45 micron infrared spectroscopy of carbon-rich proto-planetary nebulae. Astrophys. J. 535:275-292.

Hu, A., and W.W. Duley. 2008. Spectra of carbon nanoparticles: laboratory simulation of the aromatic CH emission feature at 3.29 $\mu$m. Astrophys J. 677, L153-156.

Hudgins, D.M. and L.J. Allamandola. 1999. The spacing of the interstellar 6.2 and 7.7 micron emission features as an indicator of polycyclic aromatic hydrocarbon size. Astrophys. J. Let. 513: L69-L73.

Hudgins, D.M., C.W. Bauschlicher, and L.J. Allamandola. 2005. Variations in the peak position of the 6.2 μm interstellar emission feature: a tracer of N in the interstellar polycyclic aromatic hydrocarbon population. Astrophys. J. 632:316-332.

Ibarra, J., E. Muñoz, and R. Moliner. 1996. FTIR study of the evolution of coal structure during the coalification process. Organic Geochemistry 24(6):725-735.

Jäger C., F.J. Molster, J. Dorschner, T. Henning, H. Mutschke, and L. Waters. 1998. Steps toward interstellar silicate mineralogy - IV. The crystalline revolution. Astron. Astrophys. 339(3):904-916.

Jäger, C., F. Huisken, H. Mutschke, I.L. Jansa, and Th. Henning. 2009. Formation of polycyclic aromatic hydrocarbons and carbonaceous solids in gas-phase condensation experiments. Astrophys. J. 696(1):706-712.

Jones, A.P., W.W. Duley, and D.A. Williams. 1990. The structure and evolution of hydrogenated amorphous carbon grains and mantles in the interstellar medium. QJRAS 31:567-582.

Jones, A.P. 2012a. Variations on a theme - the evolution of hydrocarbon solids. I. Compositional and spectral modelling - the eRCN and DG models. Astron. Astrophys. 540:1.

Jones, A.P. 2012b. Variations on a theme - the evolution of hydrocarbon solids. II. Optical property modelling - the optEC(s) model. Astron. Astrophys. 540:2.

Jones, A.P. 2012c. Variations on a theme - the evolution of hydrocarbon solids (corrigendum). III. Size-dependent properties - the optEC(s)(a) model. Astron. Astrophys. 545:3.

Jones, A.P., L. Fanciullo, M. Köhler, L. Verstraete, V. Guillet, M. Bocchio, and N. Ysard. 2013. The evolution of amorphous hydrocarbons in the ISM: dust modelling from a new vantage point. Astron. Astrophys. 558:62.

Kahanpää, J., K. Mattila, K. Lehtinen, C. Leinert, and D. Lemke. 2003. Unidentified infrared bands in the interstellar medium across the Galaxy. Astron. Astrophys. 405:999-1012.

Keifer, J.R., M. Novicky, M.S. Akhter, A.R. Chughtai, and D.M. Smith. 1981. The nature and reactivity of the elemental carbon (soot) surface as revealed by Fourier transform infrared (FTIR) Spectroscopy. 1981 International Conference on Fourier Transform Infrared Spectroscopy, (SPIE), p 5.

Keller, L. P., S. Bajt, G.A. Baratta, et al. 2006. Infrared spectroscopy of Comet 81P/Wild 2 samples returned by Stardust. Science, 314, 1728-1731.

Kim, S. J., A. Jung, C.K. Sim et al. 2011, Retrieval and tentative identification of the 3 μm spectral feature in Titan's haze. Planet. Space Sci. 59:699-704.

Kim, S.J., C.K. Sim, D.W. Lee, R. Courtin, J.I. Moses, Y.C. Minh. 2012, Planet. Space Sci. 65: 122-129.

Knacke, R.F. 1977. Carbonaceous compounds in interstellar dust. Nature 269:132-134.

Kroto, H.W., J.R. Heath, S.C. Obrien, R.F. Curl, and R.E. Smalley. 1985. $C_{60}$: Buckminsterfullerene. Nature 318:162-163.

Kvenvolden K., J. Lawless, K. Pering, et al. 1970. Evidence for Extraterrestrial Amino-acids and Hydrocarbons in the Murchison Meteorite. Nature 228:923-926.

Kwok, S. 1993. Proto-planetary nebulae. Annu. Rev. Astron. Astrophys. 31:63-92.

Kwok, S. 2000. Origin and Evolution of Planetary Nebulae. Cambridge University Press.

Kwok, S. 2004. The synthesis of organic and inorganic compounds in evolved stars. Nature 430:985-991.

Kwok, S. 2011. Organic Matter in the Universe. Wiley.

Kwok, S. 2017. Abiotic synthesis of complex organics in the Universe. Nature Astronomy 1(10):642.

Kwok, S., and Y. Zhan. 2011. Mixed aromatic-aliphatic organic nanoparticles as carriers of unidentified infrared emission features. Nature 479:80-83.

Kwok, S. and Y. Zhang. 2013. Unidentified Infrared Emission Bands: PAHs or MAONs? Astrophys. J. **771**, 5

Kwok, S., K. Volk, and W.P. Bidelman. 1997. Classification and Identification of IRAS Sources with Low-Resolution Spectra. Astrophys J Suppl Ser 112:557-584.

Kwok, S., K. Volk, and P. Bernath. 2001. On the origin of infrared plateau features in proto-planetary nebulae. Astrophys. J. 554: L87-L90.

Kwok, S., K.M. Volk, and B.J. Hrivnak. 1989. A 21 micron emission feature in four proto-planetary nebulae. Astrophys. J. 345: L51-L54.

Léger, A. and J.L. Puget. 1984. Identification of the 'unidentified' IR emission features of interstellar dust? Astron. Astrophys. 137: L5-L8.

Mennella, V., J.R. Brucato, L. Colangeli, and P. Palumbo. 1999. Activation of the 3.4 micron band in carbon grains by exposure to atomic hydrogen. Astrophys. J. 524: L71-L74.

Mennella, V., G.A. Baratta, A. Esposito, G. Ferini, and Y.J. Pendleton. 2003. The effects of ion irradiation on the evolution of the carrier of the 3.4 micron interstellar absorption band. Astrophys. J. 587:727-738.

Meringer, M. and H.J. Cleaves. 2017. Exploring astrobiology using "in silico" molecular structure generation. Philosophical Transactions of the Royal Society A: Mathematical, Physical and Engineering Sciences 375 (2109).

Miller, S. L. 1953. A production of amino acids under possible primitive earth conditions. Science, 117: 528-529.

Nagy, B., D.J. Hennessy, and W.G. Meinschein. 1961. Mass spectroscopic analysis of the Orgueil meteorite: evidence for biogenic hydrocarbons, Annals of the NY Academy of Sciences 93, 27-35

Ney, E.P. and B.F. Hatfield. 1978. The isothermal dust condensation of Nova Vulpeculae 1976. Astrophys. J. 219: L111-L115.

Oparin, A.I. 1938. The Origin of Life. MacMillan, New York.

Painter, P.C., R.W. Snyder, M. Starsinic et al. 1981. Concerning the application of FT-IR to the study of coal: a critical assessment of band assignments and the application of spectral analysis programs. Appl. Spectrosc. 35(5):475-485.

Papoular, R. 2001. The use of kerogen data in understanding the properties and evolution of interstellar carbonaceous dust. Astron. Astrophys. 378:597-607.

Papoular, R., J. Conrad, M. Giuliano, J. Kister, and G. Mille. 1989. A coal model for the carriers of the unidentified IR bands. Astron. Astrophys. 217:204-208.

Pino, T., E. Dartois, A.-T. Cao et al. 2008. The 6.2 μm band position in laboratory and astrophysical spectra: a tracer of the aliphatic to aromatic evolution of interstellar carbonaceous dust. Astron. Astrophys. 490:665-672.

Pizzarello, S. and E. Shock. 2017. Carbonaceous chondrite meteorites: the chronicle of a potential evolutionary path between stars and life. Origins of Life and Evolution of Biospheres 47(3):249-260.

Pizzarello, S., L.B. Williams, J. Lehman, G.P. Holland, and J.L. Yarger. 2011. Abundant ammonia in primitive asteroids and the case for a possible exobiology. Proceedings of the National Academy of Science 108:4303-4306.

Posch, T., F. Kerschbaum, H. Mutschke. et al. 1999, On the origin of the 13 µm feature. A study of ISO-SWS spectra of oxygen-rich AGB stars. Astron. Astrophys. 352:609-618.

Posch, T., F. Kerschbaum, H. Mutschke, J. Dorschner, and C. Jäger. 2002. On the origin of the 19.5 µm feature. Identifying circumstellar Mg-Fe-oxides. Astron. Astrophys. 393: L7-L10.

Puetter, R.C., R.W. Russell, S.P. Willner, and B.T. Soifer. 1979. Spectrophotometry of compact H II regions from 4 to 8 microns. Astrophys. J. 228:118-122.

Robertson, J. and E.P. O'Reilly 1987. Electronic and atomic structure of amorphous carbon. Phys. Rev. B 35(6):2946-2957.

Russell, R.W., B.T. Soifer, B.T., and S.P. Willner. 1977. The 4 to 8 micron spectrum of NGC 7027. Astrophys. J. 217: L149-L153.

Sadjadi, S., Y. Zhang, and S. Kwok. 2015a. A theoretical study on the vibrational spectra of polycyclic aromatic hydrocarbon molecules with aliphatic sidegroups. Astrophys. J. 801:34.

Sadjadi, S., Y. Zhang, and S. Kwok. 2015b. On the origin of the 11.3 micron unidentified infrared emission feature. Astrophys. J. 807:95.

Sakata, A., S. Wada, T. Onaka, and A.T. Tokunaga. 1987. Infrared spectrum of quenched carbonaceous composite (QCC). II - A new identification of the 7.7 and 8.6 micron unidentified infrared emission bands. Astrophys. J. 320: L63-L67.

Sakata, A., S. Wada, T. Onaka, and A.T. Tokunaga. 1990. Quenched carbonaceous composite. III - Comparison to the 3.29 micron interstellar emission feature. Astrophys. J. 353:543-548.

Salama, F., G.A. Galazutdinov, J. Krełowski, et al. 2011. Polycyclic Aromatic Hydrocarbons and the Diffuse Interstellar Bands: A Survey. Astrophys. J. 728:154.

Sarre, P. J. 2006. The diffuse interstellar bands: A major problem in astronomical spectroscopy. J. Mol. Spectrosc. **238**, 1-10.

Scott, A. and W.W. Duley. 1996. The Decomposition of hydrogenated amorphous carbon: a connection with polycyclic aromatic hydrocarbon molecules. Astrophys. J. 472: L123-L125.

Sellgren, K. 1984. The near-infrared continuum emission of visual reflection nebulae. Astrophys. J. 277:623-633.

Sellgren, K., M.W. Werner, J.G. Ingalls et al. 2010. $C_{60}$ in reflection nebulae. Astrophys. J. 722: L54-L57.

Schmitt-Kopplin, P., Z. Gabelica, R.D. Gougeon et al. 2010. High molecular diversity of extraterrestrial organic matter in Murchison meteorite revealed 40 years after its fall. Proceedings of the National Academy of Science 107:2763-2768.

Sherwood Lollar, B., T.D. Westgate, J.A. Ward, G.F. Slater, and G. Lacrampe-Couloume. 2002. Abiogenic formation of alkanes in the Earth's crust as a minor source for global hydrocarbon reservoirs. Nature 416, 522-524.

Tielens, A.G.G.M. 2008. Interstellar polycyclic Aromatic Hydrocarbon Molecules. Annu. Rev. Astron. Astrophys. 46:289-337.

Uchida, K.I., K. Sellgren, M.W. Werner, and M.L. Houdashelt. 2000. Infrared Space Observatory mid-infrared spectra of reflection nebulae. Astrophys. J. 530(2):817-833.

van Diedenhoven, B., E. Peeters, C. Van Kerckhoven et al. 2004. The profiles of the 3-12 micron polycyclic aromatic hydrocarbon features. Astrophys. J. 611:928-939.

Van Winckel, H., M. Cohen, and T.R. Gull. 2002. The ERE of the Red Rectangle revisited.

Astron. Astrophys. 390: 147-154.

Volk, K., S. Kwok, S., and B.J. Hrivnak. 1999. High-resolution Infrared Space Observatory spectroscopy of the unidentified 21 micron feature. Astrophys. J. 516: L99-L102.

Volk, K., G.-Z. Xiong, and S. Kwok. 2000. Infrared Space Observatory spectroscopy of extreme carbon stars. Astrophys. J. 530: 408-417.

Wagner, D.R., H. Kim, and R.J. Saykally. 2000. Peripherally hydrogenated neutral polycyclic aromatic hydrocarbons as carriers of the 3 micron interstellar infrared emission complex: results from single-photon infrared emission spectroscopy. Astrophys. J. 545:854-860.

Witt, A.N., K.D. Gordon, and D.G. Furton. 1998. Silicon nanoparticles: source of extended red emission? Astrophys. J. 501: L111-L115.

Woolf, N.J. and E.P. Ney. 1969. Circumstellar infrared emission from cool stars. Astrophys. J. 155: L181-L184.

Yabuta, H., L.B. Williams, G.D. Cody, G.D., C.M.O.D. Alexander, and S. Pizzarello. 2007. The insoluble carbonaceous material of CM chondrites: A possible source of discrete organic compounds under hydrothermal conditions. Meteoritics and Planetary Science 42:37-48.

Zhang, Y. and S. Kwok. 2011. Detection of $C_{60}$ in the protoplanetary nebula IRAS 01005+7910. Astrophys. J. 730:126.

Zhang, Y. and S. Kwok. 2013. On the detections of $C_{60}$ and derivatives in circumstellar environments. Earth, Planets, and Space 65:1069-1081.

Zhang, Y. and S. Kwok. 2015. On the viability of the PAH model as an explanation of the unidentified infrared emission features. Astrophys. J. 798:37.

Zhang, Y., S. Sadjadi, C.-H. Hsia, and S. Kwok. 2017. Search for hydrogenated $C_{60}$ (fulleranes) in circumstellar envelopes. Astrophys. J. 845: 76.

Zinner, E. 1998. Stellar nucleosynthesis and the isotopic composition of presolar grains from primitive meteorites. Annu. Rev. Earth Planet. Sci. 26:147-188.

Ziurys, L.M. 2006, Interstellar chemistry special feature: the chemistry in circumstellar envelopes of evolved stars: following the origin of the elements to the origin of life, Proceedings of the National Academy of Science, **103**, 12274-12279.

Ziurys, L.M., D.T. Halfen, W. Geppert, Y. Aikawa. 2016. Following the interstellar history of carbon: from the interiors of stars to the surfaces of planets. Astrobiology 16:997-1012.